\documentclass{ws-p8-50x6-00}

\usepackage{bm}
\def\vec#1{\bm{#1}}

\def\YF{\em Yad. Fiz.}

\def\MPLA{{\em Mod. Phys. Lett.} A}
\def\FBS{\em Few-Body Syst.}

\def\beq{\begin{equation}}
\def\eeq{\end{equation}}
\def\beqn{\begin{eqnarray}}
\def\eeqn{\end{eqnarray}}
\def\nn{\nonumber\\ }

\begin{document}

\title{The GDH sum rule for the $\Delta$ isobar:\\ A possible anomaly?}
\author{A. I. L'vov}

\address{Lebedev Physical Institute, Leninsky Prospect 53,
Moscow 117924, Russia
\\E-mail: lvov@x4u.lebedev.ru}

\maketitle

\abstracts{
The GDH sum rule is discussed for the $\Delta(1232)$ resonance.  It is
shown that apart from ordinary excitations to higher-energy states, the
sum rule contains a large negative contribution due to de-excitation
into the nucleon state. Therefore, a fulfillment of the sum rule
assumes a strong coupling of $\Delta^+$ and $\Delta^0$ to resonances of
spin $\ge\frac52$.  Calculations performed in quark models suggest that
$D_{15}(1675)$ may be such a resonance. However, its strength is found
to be not sufficient for bringing the GDH sum rule to a theoretically
expected positive magnitude.}

\section{Introduction}

The Gerasimov--Drell--Hearn sum rule puts a nontrivial constraint on a
spin dependence of the total photoabsorption cross section by hadrons.
For a target $B$ of spin $s$, the electric charge $eZ$ ($e^2=1/137$),
and the mass $m$ it states~\cite{gera65}
\beq
    I^{\rm GDH} \equiv \int_{\rm thr}^\infty \Big(
    \sigma_{1+s}(\omega) - \sigma_{1-s}(\omega) \Big)\,
    \frac{d\omega}{\omega} = 4\pi^2 s
         \Big(\frac{\mu}{s} - \frac{eZ}{m}\Big)^2,
\label{GDH}
\eeq
where $\mu$ is the magnetic moment of $B$ and $\mu-eZs/m$ is the anomalous
magnetic moment (a.m.m.).  The cross sections $\sigma_\lambda(\omega)$
refer to absorption of a circularly polarized photon by the target with
spin $s$ parallel or antiparallel to the photon helicity, so that
$\lambda=1\pm s$ is the net helicity.

The only strong assumption needed for the validity of the GDH relation
(\ref{GDH}), viz.\ that spin dependence of the Compton forward
scattering amplitude vanishes at high energies, seems to agree with
QCD.\cite{bass97} Recent measurements of $\sigma_\lambda(\omega)$ off
the proton at MAMI~\cite{mami00} confirm that the GDH integral is
indeed equal to its theoretical value given by the a.m.m.\ of the
proton.

We concentrate here on a contribution to the GDH integral from baryon
resonances $B^*$. In the zero-width approximation and in terms of
standard photocouplings,
\beq
    A_\lambda = \sqrt{\frac{\pi}{\omega^*}}\,
        \langle B^*(+{\textstyle\frac12}\vec k, \lambda) | J_+(0) |
       B(-{\textstyle\frac12}\vec k, \lambda-1) \rangle,
    \quad \omega^* \equiv \frac{|{m^*}^2-m^2|}{2m^*}
\label{A-lambda}
\eeq
($J_\pm = J_x \pm iJ_y$ is the electromagnetic current), this
contribution reads
\beq
    I_{\rm res}^{\rm GDH} = \sum_{B^* \ne B} \frac{2\pi}{\omega}
      \Big( |A_{1+s}|^2 - |A_{1-s}|^2 \Big),
    \quad \omega \equiv \frac{|{m^*}^2-m^2|}{2m} \, .
\label{GDH-resonance}
\eeq
Both the experimental data~\cite{mami00} and theoretical arguments of
quark models~\cite{li93,desa94} and large-$N_c$ QCD~\cite{cohen00}
suggest that the GDH integral for the nucleon is dominated by energies
in the $\Delta$-resonance region. The magnetic $\gamma N\Delta$
transition and $s$-wave pion photoproduction dominate the integral,
whereas a contribution of resonances of $N\ge 1$ oscillator bands is
relatively small. For other targets situation is, however, different,
and a fulfillment of the GDH sum rule requires $N=1$ resonance
contributions, as it will be explained below.

\section{M1-part of the GDH integral in NQM}

It is known that magnetic spin-transitions like $\gamma N\Delta$ are
generally insufficient for saturation of the GDH sum rule.  For
example, in the case of a weakly-bound system of a few nonrelativistic
$s$-wave quarks with the operator of the total magnetic moment
$\vec M = \sum_q (e_q\vec\sigma_q /2m_q)$, the (unretarded) magnetic
M1-contribution to Eq.\ (\ref{GDH-resonance}) takes the closure
form~\cite{gera65}
\beqn
    I_{\rm M1}^{\rm GDH} &=& 2\pi^2 \sum_{B^* \ne B} \Big\{
       |\langle B^*(s_z{+}1)| M_+ | B(s_z)\rangle|^2
     - |\langle B^*(s_z{-}1)| M_- | B(s_z)\rangle|^2 \Big\}
\nn &=&
    2\pi^2\, \langle B(s_z)| [ M_-, M_+ ] | B(s_z)\rangle
  - 2\pi^2\, \langle B(s_z)| [ \mu_-, \mu_+ ] | B(s_z)\rangle.
\label{GDH-M1-init}
\eeqn
Here the magnetic moment $\vec\mu$ of the state $B$ appears as
\beq
  \langle B(s_z')| \vec\mu | B(s_z)\rangle =
  \langle B(s_z')| \vec M  | B(s_z)\rangle, \quad
      \frac{\mu}{s} = \sum_q \frac{e_q}{m_q}\, P_q\, ,
\eeq
where quantities $P_q =\langle s_{qz}\rangle/s_z$, $\sum_q P_q =1$,
are fractions of the total spin $s_z=s$ carried by different quarks $q$.
In this notation Eq.\ (\ref{GDH-M1-init}) becomes
\beq
    I_{\rm M1}^{\rm GDH} = 4\pi^2 s \Big[
         \Big(\frac{\mu}{s}\Big)^2
       - \sum_q \Big(\frac{e_q}{m_q}\Big)^2 P_q \Big].
\label{GDH-M1}
\eeq

Generally, $I_{\rm M1}^{\rm GDH}$ is not equal to the r.h.s.\ of Eq.\
(\ref{GDH}).  For SU(3)-octet states like the proton ($p=uud$), for
which $P_1=P_2=\frac23$, $P_3=-\frac13$ and $e_1=e_2$, Eqs.\
(\ref{GDH}) and (\ref{GDH-M1}) do coincide, provided all masses of
quarks are the same and $m=3m_q$.  These states include also the
neutron and the strange baryons $\Sigma^\pm$, $\Xi^0$, $\Xi^-$.
However, in the case of $\Sigma^0$ and $\Lambda$ we have
$I_{\rm M1}^{\rm GDH} < I^{\rm GDH}$.  For SU(3)-decuplet states all
$P_q=\frac13$, so that the a.m.m.\ $\mu - eZs/m=0$ (in the limit of
$m=3m_q$) and therefore $I^{\rm GDH}=0$.  Meanwhile
$I_{\rm M1}^{\rm GDH}<0$ whenever there are quarks of different
electric charges in the baryon $B$.  This is the case for those members
of the decuplet which contain exactly one or two $u$-quarks and thus
have the electric charge 0 or $+1$ (e.g., $\Delta^+$ and $\Delta^0$).


\section{M1-photon scattering in NQM}

Since the a.m.m.\ of $\Delta$ is zero (at least in the NQM), a naive
understanding of the GDH integral (\ref{GDH}) as an integral over
positive $\omega$ may suggest that (spin-dependent part of) the
photoabsorption cross section off $\Delta$ is small.  This, however, is
not true since we have seen that the GDH M1-contribution
$I_{\rm M1}^{\rm GDH}$ for $\Delta^+$ or $\Delta^0$ is not zero and as
large as $-300~\mu$b.  This piece of $I^{\rm GDH}$ emerges owing to
$\Delta\to N$ transition which is de-excitation and which corresponds
to a negative-energy part of the GDH integral. In order to have
$I^{\rm GDH}=0$, a big positive contribution must also exist which can
only come from resonances of spin $\ge\frac52$.  Therefore, the GDH sum
rule implies that both $\Delta^+$ and $\Delta^0$ must have strong
electromagnetic transitions into resonances of higher spins.

In view of importance of such a conclusion, we want to argue more that
the negative de-excitation contribution is indeed a part of the GDH
integral.  Let us write the forward scattering amplitude of a photon
with the helicity $+1$ in the above-considered nonrelativistic quark
model.  Neglecting the retardation and recoil and keeping only M1 terms
in the electromagnetic current, we have
\beq
     T_{\rm M1}(\omega) = \frac{\omega^2}{2} \sum_n \Bigg(
     \frac{\langle B| M_- | n \rangle \, \langle n| M_+ | B \rangle}
          {E_n - E_B - \omega - i\epsilon} +
     \frac{\langle B| M_+ | n \rangle \, \langle n| M_- | B \rangle}
          {E_n - E_B + \omega - i\epsilon} \Bigg),
\label{T-M1}
\eeq
where the sum is taken over {\em all} possible intermediate states $n$.
At low energies $\omega$, the amplitude $T_{\rm M1}(\omega)$ is
dominated by intermediate states $|n\rangle$ of the energy $E_n=E_B$.
These states are just $|B\rangle$ with perhaps different spin
projections.  Using closure, we find that
\beq
     \omega^{-1}T_{\rm M1}(\omega) \to -\frac12 \,
         \langle B | [ \mu_-, \mu_+ ] | B \rangle
     = s \Big(\frac{\mu}{s}\Big)^2
      \quad \mbox{when $\omega\to 0$}.
\eeq
In the opposite limit of high energies we neglect $E_n-E_B$ and,
using the closure, write
\beq
     \omega^{-1}T_{\rm M1}(\omega) \to -\frac12 \,
         \langle B | [ M_-, M_+ ] | B \rangle
     = s \sum_q \Big(\frac{e_q}{m_q}\Big)^2 P_q
      \quad \mbox{when $\omega\to\infty$}.
\label{asympt}
\eeq
The conclusion is that the {\em full} GDH integral (\ref{GDH-M1-init}),
with both excited and de-excited intermediate states included, is what
determines a variation of the quantity $\omega^{-1}T_{\rm M1}(\omega)$
between low and high energies.

In the real (relativistic) world, the spin-dependent part of the full
amplitude, $T_{\rm spin}(\omega)$, is determined at low energies by the
a.m.m.\ rather than $\mu$ and $\omega^{-1} T_{\rm spin}(\omega)$
vanishes at high energies rather than goes to the constant
(\ref{asympt}).  With these changes, the full GDH integral is still
that determines a variation of $\omega^{-1}T_{\rm spin}(\omega)$
between $\omega=0$ and $\infty$ and therefore gives the r.h.s.\ of Eq.\
(\ref{GDH}).

\section{The GDH sum rule for $\Delta$ in quark models}

A challenge with evaluating the GDH sum rule for $\Delta$ is in finding
a source for a big photoabsorption cross section $\sigma_{5/2}(\omega)$
which must compensate the large negative de-excitation contribution
from the nucleon, as well as negative contributions of all resonances
of spin $\le \frac32\rule[-1ex]{0ex}{2ex}$.
Theoretical evaluations of the GDH sum rule for weakly-bound
systems~\cite{li93,desa94,kraj70} suggest that the full electromagnetic
spin-orbit interaction, including a relativistic two-body correction,
which leads to $p$-wave excitations of the system, should play an
important role here.  Using such an interaction in the framework of the
Karl--Isgur nonrelativistic quark model, we calculated photocouplings
$A_\lambda$ of the $\Delta$ resonance to all the lowest $L=1$ baryons
and found the resonance contribution (\ref{GDH-resonance}).  We have
found that the spin-orbit interaction strongly affects photocouplings
of $|N^4~P_M\rangle$ resonances (lying in the 1700 MeV mass range) and
makes $D_{15}(1675)$ a prominent mode of the $\Delta$ photoexcitation.
Still, the found strength of the $D_{15}(1675)$ contribution is not
sufficient to bring the GDH integral to a positive (even small) value
(see Table~1).

Recently, Carlson and Carone estimated photocouplings of $\Delta$ with
the $L=1$ baryons using an operator structure of the large-$N_c$ QCD
and determining unknown coefficients through {\em experimental} data on
photocouplings of the nucleon.\cite{carl98}  Depending on whether
two-body operators are included or not included into the fits, two
solutions were provided (CC-II and CC-I, respectively) which lead to
results which we shown in Table~1. Doing so, we add the contribution of
the nucleon with the {\em experimentally} known photocoupling, just to
be in line with the whole ideology of this approach.  The CC
predictions reveal a strong photocoupling of $\Delta$ with
$D_{15}(1675)$ which leads to an essential cancellation between the
$N(939)$ and $D_{15}(1675)$ contributions. Uncertainties in
photocouplings make it difficult to predict unambiguously the GDH
integral. A clear trend, however, is that the $D_{15}(1675)$ resonance
does not yield all the needed cross section $\sigma_{5/2}$.

It would be desirable to extend the present consideration by including
soft-pion photoproduction which is known to visibly contribute to the
GDH integral for the nucleon target and which is partly responsible for
reducing the otherwise too big contribution of $\Delta$ if taken with
the {\em experimental} strength.  Then effects of the finite width of
the $\Delta$ should be taken into account as well. We might anticipate
that essentially $s$-wave pion photoproduction off the $\Delta$ does
not contribute much to $\sigma_{5/2}$ but does contribute to
$\sigma_{-1/2}$ owing to the reaction $\gamma\Delta\to\pi N$. It thus
can further increase the gap between the negative resonance
contribution $I_{\rm res}^{\rm GDH}$ and the positive theoretical value
of $I^{\rm GDH}$.

\section*{Acknowledgments}

I appreciate a useful discussion with S.B.\ Gerasimov. The hospitality
and support of the Institut f\"ur Kernphysik at University of Mainz,
where a part of this work was done, is greatly acknowledged.

\begin{table}[t]
\caption{Contributions of the nucleon and $[70,1^-]$ resonances to the 
GDH integral $I^{\rm GDH}$ (in $\mu$b) for the $\Delta^+$ and 
$\Delta^{++}$ targets. NQM and NQM$+$so label the Karl--Isgur 
nonrelativistic quark model respectively without and with one- and 
two-body spin-orbit electromagnetic interactions. See other notations 
in the text. In the case of $\Delta^{++}$, results of NQM and NQM$+$so 
are identical, and almost so are results of CC-I and CC-II.
\label{table}}
\begin{center}
\footnotesize
\begin{tabular}{|c|r|r|r|r|r|r|}
\hline
   & \multicolumn{4}{|c }{$B=\Delta^+$} &
     \multicolumn{2}{|c|}{$B=\Delta^{++}$\rule[-1.3ex]{0ex}{4ex}} \\
\cline{2-7}
   $B^*$           &  NQM & NQM$+$so & CC-I &  CC-II
                          & NQM      & CC-I\rule[-1.3ex]{0ex}{4ex}\\
\hline
   $N(939)$        &$-270$&  $-270$  &$-468$& $-468$\rule[-1.3ex]{0ex}{4ex}
                          &          &       \\
\hline
   $S_{11}(1535)$  &$ -56$&  $ -40$  &$ -84$& $-149$\rule{0ex}{2.5ex}
                          &          &       \\
   $D_{13}(1520)$  &$ -28$&  $  -3$  &$ -29$& $ -25$
                          &          &       \\
   $S_{11}(1650)$  &$-192$&  $ -66$  &$ -20$& $ -57$
                          &          &       \\
   $D_{13}(1700)$  &$ -33$&  $ -18$  &$  -8$& $ -35$
                          &          &       \\
   $D_{15}(1675)$  &$ 189$&  $ 332$  &$ 309$& $ 532$
                          &          &       \\
   $S_{31}(1620)$  &$ -12$&  $ -12$  &$ -10$& $ -10$
                          &  $ -47$  &$ -38$ \\
   $D_{33}(1700)$  &$ -29$&  $ -29$  &$ -13$& $ -13$\rule[-1.3ex]{0ex}{2ex}
                          &  $-116$  &$ -51$ \\
\hline
   total ($\mu$b)  &$-431$&  $-105$  &$-323$& $-225$\rule[-1.3ex]{0ex}{4ex}
                          &  $-163$  &$ -90$ \\
\hline
\end{tabular}
\end{center}
\end{table}

\end{document}